\documentclass[10pt,conference]{IEEEtran}
\IEEEoverridecommandlockouts
\usepackage{cite}
\usepackage[switch]{lineno}
\usepackage{amsmath,amssymb,amsfonts}
\usepackage{algorithmic}
\usepackage{graphicx}
\usepackage{textcomp}
\usepackage{xcolor}
\usepackage{url}

\def\BibTeX{{\rm B\kern-.05em{\sc i\kern-.025em b}\kern-.08em
    T\kern-.1667em\lower.7ex\hbox{E}\kern-.125emX}}
\begin{document}
\title{Story Point Effort Estimation by Text Level Graph Neural Network}
\author{\IEEEauthorblockN{Hung Phan}
\IEEEauthorblockA{\textit{Department of Computer Science} \\
\textit{Iowa State University}\\
Ames, Iowa, USA \\
hungphd@iastate.edu}
\and
\IEEEauthorblockN{Ali Jannesari}
\IEEEauthorblockA{\textit{Department of Computer Science} \\
\textit{Iowa State University}\\
Ames, Iowa, USA \\
jannesar@iastate.edu}
}

\maketitle

\begin{abstract}
Estimating the software projects’ efforts developed by agile methods is important for project managers or technical leads. It provides a summary as a first view of how many hours and developers are required to complete the tasks. There are research works on automatic predicting the software efforts, including Term Frequency - Inverse Document Frequency (TFIDF) as the traditional approach for this problem. Graph Neural Network is a new approach that has been applied in Natural Language Processing for text classification. The advantages of Graph Neural Network are based on the ability to learn information via graph data structure, which has more representations such as the relationships between words compared to approaches of vectorizing sequence of words. In this paper, we show the potential and possible challenges of Graph Neural Network text classification in story point level estimation. By the experiments, we show that the GNN Text Level Classification can achieve as high accuracy as about 80\% for story points level classification, which is comparable to the traditional approach. We also analyze the GNN approach and point out several current disadvantages that the GNN approach can improve for this problem or other problems in software engineering.  
\end{abstract}


\section{Introduction}

With the importance of story point effort estimation, there are researches on how to apply the technique to automatically predict the story point of a software task/ issue. To our knowledge, one of the well-known datasets of software effort estimation is the dataset published by Choetkiertikul et al \cite{004_DeepLearningSPEstimation}. Machine learning-based approaches have been applied to this dataset for story point estimations by \cite{004_DeepLearningSPEstimation}. The approaches range from the classical approach from Bag of Word (BOW) \cite{016_TFIDFBook} to modern deep learning approaches from Recurrent Highway Network (RHN) and Long Short Term Memory (LSTM). In general, these approaches are formalized as machine learning regression problems.


The input of software effort estimation for each software issue is the title and description of issues. The output is the number of story points that usually ranged from 1 to 100. According to \cite{007_FibonacciAgile}, the value of story point should be in the modified version of Fibonacci numbers to ensure that the different story points are not too close to each other. Since then, the distribution of story points is range from 1 to 100 but not in continuous uniform distribution. The input of this problem can be considered as a type of software documentation that is used for requirement collection.  

 Since Graph Neural Network (GNN) had been applied in several research works in SE \cite{005_allamanisgraph,006_GNNPaper2}, it is a new area to apply GNN techniques on software requirement datasets. In this work, we want to study the potential and challenges when adapting a GNN model in text classification for story points categorization. We split the story points label in the dataset \cite{004_DeepLearningSPEstimation} into four levels. Next, we build the training model using a technique called Text Level GNN \cite{008_TextLevelGNN}. This training model is used for story points level prediction. We published our code and dataset at this site\footnote{https://github.com/pdhung3012/StoryPointEstimation-TextLevelGNN}.

\section{Related Work}
In the problem that we considered as text classification in SE, there are several approaches to applying GNN for this problem. Compared to other deep neural network based model such as in \cite{10.1145/3416506.3423576} and \cite{DBLP:conf/kbse/PhanS021}, GNN has an advantages of learning from graphs. The first approach is proposed in \cite{009_GraphGCN1}. In this work, there are two types of nodes constructed from the graph. They are nodes as documents and nodes as words inside documents. The second approach is called text level graph in NLP \cite{008_TextLevelGNN}, which worked by extracting the graph for each input text or document. Compared to \cite{009_GraphGCN1}, \cite{008_TextLevelGNN} is considered as an improvement approach since it can construct a graph for each document that consumes less memory and be feasible to interact with unseen test data. In our work, we inherit the model from \cite{008_TextLevelGNN} for our problem. 


\section{Approach}
\begin{figure*}
        \center{\includegraphics[width=0.9\linewidth]
        {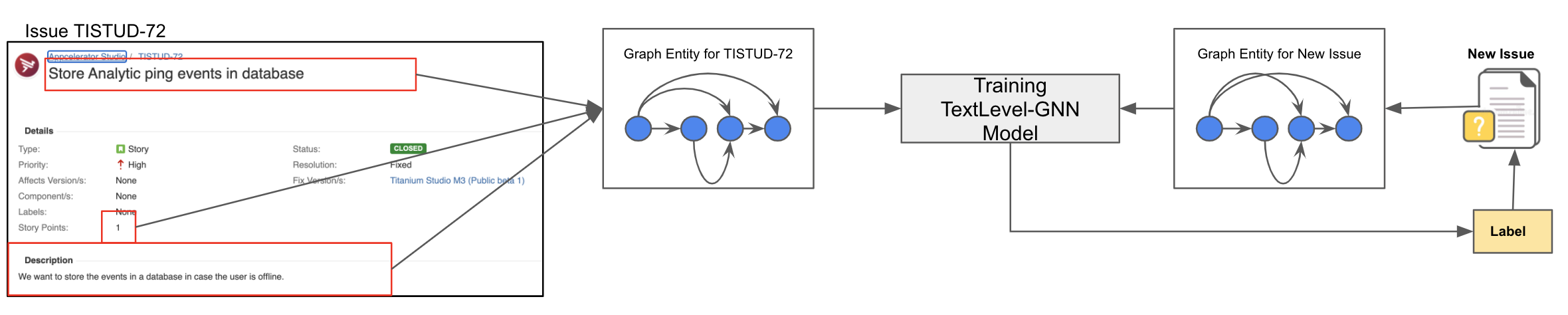}}
        \caption{Overview Architecture of TextGraph GNN model}
        \label{figOverview} 
\end{figure*}

In this project, we try to solve the story point estimation problem by inheriting the Text Graph Neural Network proposed by \cite{008_TextLevelGNN}. We call our solution as TextLevelGNN-StoryPointEstimator.The overview architecture of our approach can be shown in Figure \ref{figOverview}.

The target object given to each issue is the level of story points. According to the distribution of software effort value, we divide the level of story points into 4 levels. They are small (story point is in range [1,5] ), medium ([6,15]), large ([16,40]) and huge ([41,100]). This distribution can be summarized in Table \ref{tblDistribution}.The process of issue extraction can be illustrated as follows. First, a text combination of title and description will be concatenated. Second, a graph will be constructed for each issue, which reflects the relationship between each word and its adjacent words. Finally, a GNN model is used to train a model which can infer the software effort levels. Next, we discuss important modules provided in our approach.

\subsection{Graph Construction}
Each issue contains two textual information: the issue's title and the issue's description. We provide a combination of title and description as the textual representation of a software issue. There are several types of graph construction from the text. For this problem, we inherit the graph construction from \cite{008_TextLevelGNN}. The idea of graph generation is to make an edge from a word to an offset of adjacent words in a sequence. There can be the case that there are edges that appeared multiple times between two specific words. Those cases will be reflected by the edge weight that highlights the popularity of co-occurrence between words. Edges that appeared rarely between words are considered as public edges. The offset of neighbor words between a specific word can be called the sliding window offset $w$. The sliding window offset is larger, the number of edges is increased, causing the increment of the complexity of the graph.


\subsection{Graph Neural Network}
One important element to construct a graph neural network is the process of passing information between nodes and edges inside the graph. This mechanism is called Message Passing Mechanism (MPM).  In this problem, TextLevelGNN \cite{008_TextLevelGNN} learns the representation for each node by its neighbor and then combines the representations of all nodes/ words to predict the label. The initial vector for each node is retrieved by the vocabulary of Glove vector representation. \cite{013_Glove}. The combination mechanism to predict the label can be done by the softmax and Relu functions \cite{008_TextLevelGNN}.

\section{Evaluation}
\begin{figure}
        \center{\includegraphics[width=0.75\linewidth]
        {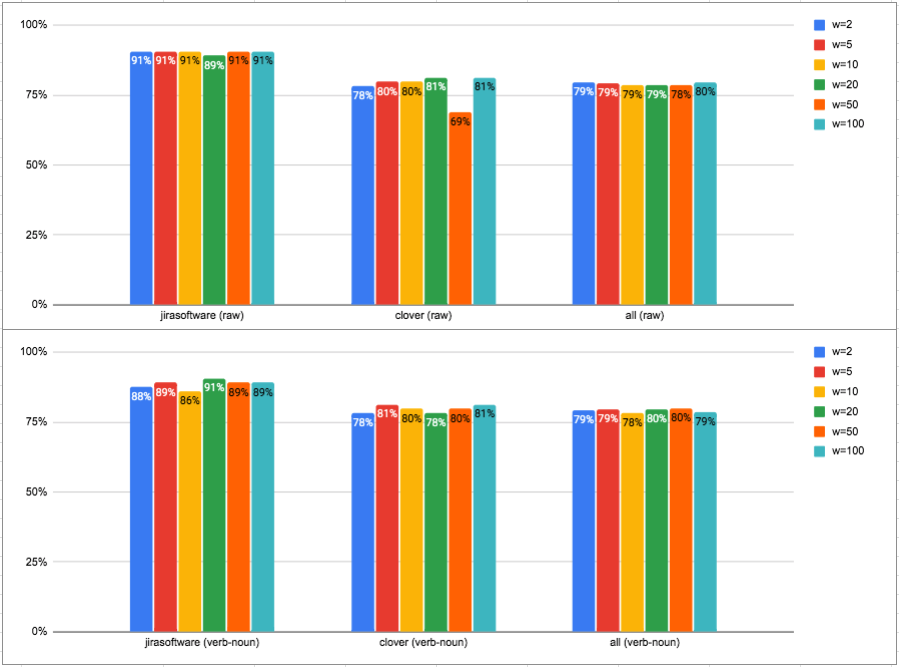}}
        \caption{Sensitivity experiment on projects jirasoftware, clover and average accuracy of 16 projects with window size as $[2,5,10,20,50,100]$ and text format as $[raw, filter verb and noun]$}
        \label{figSensitivityWindow} 
\end{figure}
\subsection{Dataset and Configurations} We use the dataset collected from \cite{004_DeepLearningSPEstimation}. There are 16 software projects developed by Agile development, which contains over 23000 software issues and actual story points. Similar to \cite{004_DeepLearningSPEstimation}, we split the dataset as 80\% for training and validation, 20\% for testing. We use the same configuration with prior work \cite{008_TextLevelGNN}, with batch size as 32, drop out probability as 0.5, sliding window offset as 20, and the default Glove representation pre-trained model with 300-dimensional vectors for each word in English. The distribution of the level of story points and numbers of issues per each level are shown in Table \ref{tblDistribution}. We select the range of story point levels based on the median of story points for each project on dataset \cite{004_DeepLearningSPEstimation}.

We compare the GNN approach with the traditional approach of story point level classification using TFIDF vectorization and Random Forest for classification. For TFIDF vectorization, we use the n-gram as the combination from 1-gram to 4-grams and use the library TfIdfVectorizer in Python \cite{015_TFIDFVectorizer}. For RandomForest (RF) classification, we use its implementation from Scikit-learn library \cite{023_Scikitlearn}.  We call the traditional approach TFIDF-RF. We aim to answer the following research questions (RQs):

\begin{enumerate}
    \item \textbf{RQ 1}. Accuracy of TextLevel GNN classification from raw text of issues' titles and description.
    \item \textbf{RQ 2}. Analysis on ability of TextLevel GNN for regression for story points estimation.
    \item \textbf{RQ 3}. Analysis on abilities of optimization in running time and graph complexity for TextLevel GNN.
\end{enumerate}

\begin{table}
\tiny
\centering
\caption{Distribution of Story Point Level in dataset \cite{004_DeepLearningSPEstimation}}
\begin{tabular}{|l|l|r|}
\hline
\multicolumn{1}{|c|}{\textbf{Level}} & \multicolumn{1}{c|}{\textbf{SP Value}} & \multicolumn{1}{c|}{\textbf{Size}} \\ \hline
Small                                & {[}1,5{]}                              & 16759                              \\ \hline
Medium                               & {[}6,15{]}                             & 5173                               \\ \hline
Large                                & {[}16,40{]}                            & 1085                               \\ \hline
Huge                                 & \textgreater{}40                       & 296                                \\ \hline
\end{tabular}
\label{tblDistribution}
\end{table}

\subsection{RQ 1. Results on TextLevel GNN Story Point Level Classification}

The result of TextLevelGNN is shown in Table \ref{tblAccuracy}. From the results on 16 projects, we achieve an accuracy as close as the traditional approach for classification. We achieve the highest accuracy on the bamboo project as 100\% while the lowest accuracy is the aptanastudio project as over 41\%. We achieve high accuracy as TFIDF and Random Forest in almost all projects. The project that TFIDF and RF outperform us most are mule as about 9\% while we achieve significantly higher accuracy on the talendataquality by 3\%. We analyze the data inside 16 projects. We see that the mule and talenddataquality provide challenges with low accuracy in story point regression in prior work \cite{004_DeepLearningSPEstimation}, which our experiment has consistent results in our classification problem.

\begin{table}
\tiny
\centering
\caption{Accuracy of Text Level GNN compared to. TFIDF-RF in Story Point Level Classification and Story Point Estimation}
\label{tblAccuracy}

\begin{tabular}{|r|l|r|r|r|r|}
\hline
\multicolumn{1}{|l|}{\textbf{}}   & \textbf{Metric}                        & \multicolumn{2}{l|}{\textbf{Classify Acc}}                                                                     & \multicolumn{2}{l|}{\textbf{Reg MAE}}                                                                          \\ \hline
\multicolumn{1}{|c|}{\textbf{No}} & \multicolumn{1}{c|}{\textbf{Software}} & \multicolumn{1}{c|}{\textbf{\begin{tabular}[c]{@{}c@{}}TFIDF\\ RF\end{tabular}}} & \multicolumn{1}{c|}{\textbf{GNN}} & \multicolumn{1}{c|}{\textbf{\begin{tabular}[c]{@{}c@{}}TFIDF\\ RFR\end{tabular}}} & \multicolumn{1}{c|}{\textbf{GNN}} \\ \hline
1                                 & appceleratorstudio & 78.08\%                                                                          & 77.78\%                           & 1.66                                                                              & 2.67                              \\ \hline
2                                 & aptanastudio       & 46.99\%                                                                          & 41.41\%                           & 3.50                                                                              & 5.16                              \\ \hline
3                                 & bamboo             & 100.00\%                                                                         & 100.00\%                          & 0.88                                                                              & 0.00                              \\ \hline
4                                 & clover             & 83.12\%                                                                          & 81.25\%                           & 4.04                                                                              & 1.22                              \\ \hline
5                                 & datamanagement     & 61.67\%                                                                          & 56.36\%                           & 6.90                                                                              & 13.40                             \\ \hline
6                                 & duracloud          & 99.25\%                                                                          & 99.22\%                           & 0.91                                                                              & 0.07                              \\ \hline
7                                 & jirasoftware       & 91.55\%                                                                          & 89.06\%                           & 1.84                                                                              & 1.31                              \\ \hline
8                                 & mesos              & 97.02\%                                                                          & 96.88\%                           & 1.30                                                                              & 0.28                              \\ \hline
9                                 & moodle             & 59.83\%                                                                          & 56.25\%                           & 11.17                                                                             & 25.16                             \\ \hline
10                                & mule               & 74.16\%                                                                          & 66.41\%                           & 2.46                                                                              & 4.03                              \\ \hline
11                                & mulestudio         & 54.42\%                                                                          & 52.34\%                           & 3.92                                                                              & 5.84                              \\ \hline
12                                & springxd           & 87.96\%                                                                          & 88.07\%                           & 1.90                                                                              & 1.08                              \\ \hline
13                                & talenddataquality  & 80.14\%                                                                          & 83.20\%                           & 3.99                                                                              & 1.50                              \\ \hline
14                                & talendesb          & 98.28\%                                                                          & 98.44\%                           & 0.87                                                                              & 0.14                              \\ \hline
15                                & titanium           & 75.61\%                                                                          & 74.55\%                           & 2.75                                                                              & 2.30                              \\ \hline
16                                & usergrid           & 95.88\%                                                                          & 96.88\%                           & 1.25                                                                              & 0.03                              \\ \hline
\multicolumn{1}{|l|}{}            & \textbf{Average}   & \textbf{80.25\%}                                                                 & \textbf{78.63\%}                  & \textbf{3.08}                                                                     & \textbf{3.09}                     \\ \hline
\end{tabular}
\end{table}

\subsection{RQ 2. From Text Classification to Text Regression} 
The TextLevelGNN approach \cite{008_TextLevelGNN} is applicable in text or issue classification. Though the space of story point value is not in uniform distribution, the story point estimation should be considered as the text regression problem instead of classification. We study other works in Software Engineering that applied GNN. We see that most of them focus on the classification or code generation instead of regression such as \cite{005_allamanisgraph}. There are a few problems in SE that can be formalized as regression problems such as project assignment scoring \cite{020_ProgramAssignmentScoring} and Github star project prediction \cite{021_GithubStarPrediction}, however, none of them applied GNN as solutions. We studied several works on graph regression in NLP and see that most of them focus on datasets in different research areas such as image processing \cite{022_GNNOverview}. 

\textbf{Considering story points as labels for classification} A simple solution to convert from classification to regression problem is that we consider the story points themselves as labels for classification. Next, we evaluate the correctness of output by Mean Absolute Error (MAE). We compare the traditional approach TFIDF with Random Forest Regression (TFIDF-RFR). The result which is shown in Table \ref{tblAccuracy}, reveals the fact that the TextLevelGNN approach achieves higher MAE than the BOW-RFR approach. It means the TextLevelGNN didn't perform as well as the traditional approach with this simple solution for converting from classification to regression problem. In the upcoming work, we intend to change the mechanism in \cite{008_TextLevelGNN} to regression learning to support story point estimation.

\subsection{RQ 3. Optimization in Running Time and Graph Complexity} 
We see that the running time in the training of your approach is longer than the traditional work. The BOW-RF approach requires \textbf{4} minutes to complete the training step, while the TextLevelGNN requires about \textbf{25} minutes for the same tasks of training on 16 projects of \cite{004_DeepLearningSPEstimation}. The testing step achieves almost the same running time for the TFIDF-RF approach and our approach. We study the result and see that one problem due to long training time is the number of edges in our graph is usually high, which can be more than 500000 edges in complicated projects such as $memos$. In future work, we can optimize the running time by simplifying the graph in which we highlight important information. 

\begin{table}
\tiny
\centering
\caption{Number of edges of projects jirasoftware and clover with window size as $[2,5,10,20,50,100]$}
\label{tblSensitivityEdgeSize}
\begin{tabular}{|l|r|r|}
\hline
\multicolumn{1}{|c|}{\textbf{\begin{tabular}[c]{@{}c@{}}window \\ size\end{tabular}}} & \multicolumn{1}{c|}{\textbf{jirasoftware}} & \multicolumn{1}{c|}{\textbf{clover}} \\ \hline
w=2                                                                                   & 22511                                      & 48658                                \\ \hline
w=5                                                                                   & 58752                                      & 124600                               \\ \hline
w=10                                                                                  & 99626                                      & 213987                               \\ \hline
w=20                                                                                  & 151170                                     & 339562                               \\ \hline
w=50                                                                                  & 229599                                     & 568997                               \\ \hline
w=100                                                                                 & 287869                                     & 772672                               \\ \hline
\end{tabular}
\end{table}

\begin{table}
\tiny
\centering
\caption{Analysis on scale of graphs extracted from training data}
\label{tblAnalysisResult}

\begin{tabular}{|l|l|r|r|r|r|}
\hline
\multicolumn{1}{|c|}{\textbf{Pp}} & \multicolumn{1}{c|}{\textbf{Project}} & \multicolumn{1}{c|}{\textbf{Size}} & \multicolumn{1}{c|}{\textbf{Nodes}} & \multicolumn{1}{c|}{\textbf{Edges}} & \multicolumn{1}{c|}{\textbf{\begin{tabular}[c]{@{}c@{}}Train\\ Time \\ (sec)\end{tabular}}} \\ \hline
{[}3{]}                              & appceleratorstudio                    & 1868                               & 14969                               & 1218745                             & 244                                                                                            \\ \hline
{[}3{]}                              & aptanastudio                          & 530                                & 8211                                & 545099                              & 41                                                                                             \\ \hline
{[}3{]}                              & bamboo                                & 332                                & 6175                                & 297828                              & 20                                                                                             \\ \hline
{[}3{]}                              & clover                                & 245                                & 5227                                & 339562                              & 14                                                                                             \\ \hline
\textbf{{[}3{]}}                     & \textbf{datamanagement}               & \textbf{2986}                      & \textbf{18490}                      & \textbf{1618297}                    & \textbf{358}                                                                                   \\ \hline
{[}3{]}                              & duracloud                             & 425                                & 3793                                & 289685                              & 21                                                                                             \\ \hline
{[}3{]}                              & jirasoftware                          & 224                                & 2339                                & 151170                              & 9                                                                                              \\ \hline
\textbf{{[}3{]}}                     & \textbf{mesos}                        & \textbf{1075}                      & \textbf{25494}                      & \textbf{1698010}                    & \textbf{101}                                                                                   \\ \hline
{[}3{]}                              & moodle                                & 745                                & 8468                                & 696113                              & 70                                                                                             \\ \hline
{[}3{]}                              & mule                                  & 568                                & 6133                                & 437088                              & 28                                                                                             \\ \hline
{[}3{]}                              & mulestudio                            & 468                                & 4316                                & 309768                              & 21                                                                                             \\ \hline
{[}3{]}                              & springxd                              & 2256                               & 17198                               & 1194025                             & 128                                                                                            \\ \hline
{[}3{]}                              & talenddataquality                     & 883                                & 7324                                & 445830                              & 92                                                                                             \\ \hline
\textbf{{[}3{]}}                     & \textbf{talendesb}                    & \textbf{555}                       & \textbf{8766}                       & \textbf{591293}                     & \textbf{30}                                                                                    \\ \hline
{[}3{]}                              & titanium                              & 1440                               & 23989                               & 1728925                             & 304                                                                                            \\ \hline
{[}3{]}                              & usergrid                              & 308                                & 4114                                & 218491                              & 17                                                                                             \\ \hline
{[}7{]}                              & R8 dataset                            & 4937                               & 2923                                & 1380208                             & 821                                                                                            \\ \hline
\end{tabular}
\end{table}

\begin{table}
\tiny
\centering
\caption{Accuracy on simplifying the graph in verb-noun format in 16 projects}
\label{tblSimplifyTotalAccuracy}
\begin{tabular}{|r|l|r|r|}
\hline
\multicolumn{1}{|l|}{\textbf{}}   & \textbf{Text}      & \multicolumn{1}{l|}{\textbf{RawText}} & \multicolumn{1}{l|}{\textbf{\begin{tabular}[c]{@{}l@{}}Filter \\ Verb-Noun\end{tabular}}} \\ \hline
\multicolumn{1}{|l|}{\textbf{No}} & \textbf{Software}  & \multicolumn{1}{l|}{\textbf{GNN}}     & \multicolumn{1}{l|}{\textbf{GNN}}                                                         \\ \hline
1                                 & appceleratorstudio & 77.78\%                               & 77.95\%                                                                                   \\ \hline
2                                 & aptanastudio       & 41.41\%                               & 54.69\%                                                                                   \\ \hline
3                                 & bamboo             & 100.00\%                              & 100.00\%                                                                                  \\ \hline
4                                 & clover             & 81.25\%                               & 78.13\%                                                                                   \\ \hline
5                                 & datamanagement     & 56.36\%                               & 61.72\%                                                                                   \\ \hline
6                                 & duracloud          & 99.22\%                               & 99.22\%                                                                                   \\ \hline
7                                 & jirasoftware       & 89.06\%                               & 90.63\%                                                                                   \\ \hline
8                                 & mesos              & 96.88\%                               & 96.88\%                                                                                   \\ \hline
9                                 & moodle             & 56.25\%                               & 61.98\%                                                                                   \\ \hline
10                                & mule               & 66.41\%                               & 67.97\%                                                                                   \\ \hline
11                                & mulestudio         & 52.34\%                               & 50.78\%                                                                                   \\ \hline
12                                & springxd           & 88.07\%                               & 88.07\%                                                                                   \\ \hline
13                                & talenddataquality  & 83.20\%                               & 75.78\%                                                                                   \\ \hline
14                                & talendesb          & 98.44\%                               & 98.44\%                                                                                   \\ \hline
15                                & titanium           & 74.55\%                               & 75.45\%                                                                                   \\ \hline
16                                & usergrid           & 96.88\%                               & 96.88\%                                                                                   \\ \hline
\multicolumn{1}{|l|}{}            & \textbf{Average}   & \textbf{78.63\%}                      & \textbf{79.66\%}                                                                          \\ \hline
\end{tabular}
\end{table}

\begin{table}
\tiny
\centering
\caption{Accuracy on simplifying the graph in verb-noun format in datamanagement project}
\label{tblSimplifyAccuracy}
\begin{tabular}{|l|r|r|}
\hline
\multicolumn{1}{|c|}{\textbf{Metric}} & \multicolumn{1}{c|}{\textbf{\begin{tabular}[c]{@{}c@{}}Raw \\ issue\end{tabular}}} & \multicolumn{1}{c|}{\textbf{\begin{tabular}[c]{@{}c@{}}Filter \\ Verb-Noun\end{tabular}}} \\ \hline
nodes                                 & 18490                                                                              & \textbf{2986}                                                                             \\ \hline
edges                                 & 1618297                                                                            & \textbf{10309}                                                                            \\ \hline
training time                         & 358                                                                                & \textbf{209}                                                                              \\ \hline
accuracy                              & 56.36\%                                                                            & \textbf{61.72\%}                                                                          \\ \hline
\end{tabular}
\end{table}

\subsubsection{Analysis on the scale of graph.} To study how the complexity can affect the performance of the training process, we analyze the nodes, the edges, the size (number of issues for training) and compare with similar metrics on the popular natural text dataset $R8$ which was used in the work \cite{008_TextLevelGNN}. We have some observations. First, the size of 16 projects is smaller than the R8 dataset. The number of issues for each project ranges from over 200 to 3000. Second, the number of nodes in the R8 dataset is about 3000, which is lower than some projects like $datamanagement, memos$, or $talendesb$. It means that there are more new words inside each issue instead of a standard NLP text dataset. We study some issues and we see that the description of issues is usually longer than the description that the combination of title and descriptions can be more than 10 sentences. This fact of causing the risk of out-of-vocabulary words that didn't appear in the standard vector corpus Glove used in \cite{008_TextLevelGNN}. We can improve this challenge by building a new Glove representation trained from the context of issues' description. Third, the number of edges is much larger than the number of nodes. This is because the mechanism of making edges between a word and neighbor words in the offset $w=20$ causes the exponential of the edges. Similarly, the number of edges provided in the $R8$ dataset is high which is about $1.3$ million edges. In the issues dataset, the project that has the highest number of edges is $memos$ project, due to the complexity of long issues' description. The project that has the least number of edges is the $jirasoftware$ project, which can be due to the compact number of training records. Forth, for the running time, the running time for each project ranges from about 9 seconds to 370 seconds, which is comparable with the $R8$ dataset.


\subsubsection{Simplifying the graph of issues by verb and noun.} We analyze the potential of simplifying the input text as an issue by filtering its words' roles in part of speech tagging by an experiment that its results are shown in Table \ref{tblSimplifyAccuracy} and Table \ref{tblSimplifyTotalAccuracy}. In this study, we filter the words that are verbs and nouns and remove all words in other tags of the issue. We run the experiment on the $datamanagement$ project, we called this configuration $verbNounFilter$. The results show that the accuracy is \textbf{56.36\%} on this project, compared to \textbf{61.72\%} with the normal text without the filter. Moreover, the training time and the graph complexity are alleviated. Though this is a very beginning and heuristic strategy to simplify the textual content, it shows the potential of simplifying the graph that focused on important parts of speech tags. The total average accuracy on 16 projects was raised to \textbf{79.66\%}.

\subsubsection{Sensitivity on window size.} We study the impact of window size on the number of edges and the classification accuracy by the sensitivity experiment by changing window size to one of the following sizes: $[2,5,10,20,50,100]$. We evaluate projects $jirasoftware$ and $clover$. The impact of window size on the number of edges is shown in Table \ref{tblSensitivityEdgeSize}, which shows the high increase of edges when we increase the window size. The accuracy experiment is shown in Figure \ref{figSensitivityWindow}. We can see that too big a window size can cause a decrease in the accuracy, possibly due to many non-useful edges of the constructed graph.

\section{Conclusion}
In this project, we propose a new mechanism of story point level estimation. Instead of converting text sequence to vector representation for features, we apply graph as the data structure for story points level classification. By the experiment, we see that the TextLevelGNN approach achieves comparable accuracy compared to the traditional approach using TFIDF for vectorization combined with Random Forest for machine learning classification. There are rooms for improvement in future works, which is related to adjusting the classification problem to the regression problem and optimizing the running time by filtering important information of the input text for graph construction.
\bibliographystyle{unsrt}
\bibliography{refs}

\begin{thebibliography}{10}

\bibitem{004_DeepLearningSPEstimation}
Morakot Choetkiertikul, Hoa~Khanh Dam, Truyen Tran, Trang Pham, Aditya Ghose,
  and Tim Menzies.
\newblock A deep learning model for estimating story points.
\newblock {\em IEEE Transactions on Software Engineering}, 45(7):637--656,
  2019.

\bibitem{016_TFIDFBook}
Jure Leskovec, Anand Rajaraman, and Jeffrey~David Ullman.
\newblock {\em Mining of Massive Datasets}.
\newblock Cambridge University Press, USA, 2nd edition, 2014.

\bibitem{007_FibonacciAgile}
Fibonacci agile estimation.
\newblock \url{https://tinyurl.com/3cpjbnhx}.

\bibitem{005_allamanisgraph}
Miltiadis Allamanis.
\newblock Graph neural networks on program analysis.
\newblock \url{https://tinyurl.com/r2mu7y5x}.

\bibitem{006_GNNPaper2}
Miltiadis Allamanis, Marc Brockschmidt, and Mahmoud Khademi.
\newblock Learning to represent programs with graphs.
\newblock {\em CoRR}, abs/1711.00740, 2017.

\bibitem{008_TextLevelGNN}
Lianzhe Huang, Dehong Ma, Sujian Li, Xiaodong Zhang, and Houfeng Wang.
\newblock Text level graph neural network for text classification.
\newblock {\em CoRR}, abs/1910.02356, 2019.

\bibitem{10.1145/3416506.3423576}
Hung Phan and Ali Jannesari.
\newblock {\em Statistical Machine Translation Outperforms Neural Machine
  Translation in Software Engineering: Why and How}, page 3–12.
\newblock Association for Computing Machinery, New York, NY, USA, 2020.

\bibitem{DBLP:conf/kbse/PhanS021}
Hung Phan, Arushi Sharma, and Ali Jannesari.
\newblock Generating context-aware {API} calls from natural language
  description using neural embeddings and machine translation.
\newblock In {\em 36th {IEEE/ACM} International Conference on Automated
  Software Engineering, {ASE} 2021 - Workshops, Melbourne, Australia, November
  15-19, 2021}, pages 219--226. {IEEE}, 2021.

\bibitem{009_GraphGCN1}
Liang Yao, Chengsheng Mao, and Yuan Luo.
\newblock Graph convolutional networks for text classification.
\newblock {\em CoRR}, abs/1809.05679, 2018.

\bibitem{013_Glove}
Glove tutorial.
\newblock \url{https://tinyurl.com/bmb2582v}.

\bibitem{015_TFIDFVectorizer}
Tf-idf vectorizer library.
\newblock \url{https://tinyurl.com/3a6v45kv}.

\bibitem{023_Scikitlearn}
Scikit learn tutorial.
\newblock \url{https://tinyurl.com/29ebj8v9}.

\bibitem{020_ProgramAssignmentScoring}
Milena Vujo\v{s}evi\'{c}-Jani\v{c}i\'{c}, Mladen Nikoli\'{c}, Du\v{s}an
  To\v{s}i\'{c}, and Viktor Kuncak.
\newblock Software verification and graph similarity for automated evaluation
  of students' assignments.
\newblock {\em Inf. Softw. Technol.}, 55(6):1004–1016, jun 2013.

\bibitem{021_GithubStarPrediction}
Github project star prediction.
\newblock \url{https://tinyurl.com/5djdpsc8}.

\bibitem{022_GNNOverview}
Jie Zhou, Ganqu Cui, Shengding Hu, Zhengyan Zhang, Cheng Yang, Zhiyuan Liu,
  Lifeng Wang, Changcheng Li, and Maosong Sun.
\newblock Graph neural networks: A review of methods and applications, 2021.

\end{thebibliography}

\end{document}